# Field tunable three-dimensional magnetic nanotextures in cobalt-nickel nanowires

I.M Andersen[†], D. Wolf[‡], L.A. Rodriguez[§◊], A. Lubk[‡], D. Oliveros[†], C. Bran[¶], T. Niermann[♠], U.K. Rößler[‡], M. Vazquez[¶], C. Gatel[†], E. Snoeck[†]

[†]Centre d'Élaboration de Matériaux et d'Etudes Structurales -CNRS, 29 rue Jeanne Marvig, 31055 Toulouse, France
[‡]Leibniz Institute for Solid State and Materials Research, IFW Dresden, Helmholtzstraße 20, 01069 Dresden, Germany
[§]Departamento de Física – Universidad del Valle, A.A. 25360 Cali, Colombia
[◊]Centro de Excelencia en Nuevos Materiales – Universidad del Valle, A.A. 25360 Cali, Colombia
[¶]Instituto de Ciencia de Materiales de Madrid – CSIC, Sor Juana Inés de la Cruz, 3, Madrid 28049, Spain
[♠]Institut für Optik und Atomare Physik, Technische Universität Berlin, Straße des 17. Juni 135, 10623 Berlin, Germany

**ABSTRACT** Cylindrical magnetic nanowires with large transversal magnetocrystalline anisotropy have been shown to sustain non-trivial magnetic configurations resulting from the interplay of spatial confinement, exchange, and anisotropies. Exploiting these peculiar 3D spin configurations and their solitonic inhomogeneities are prospected to improve magnetization switching in future spintronics, such as power-saving magnetic memory and logic applications. Here we employ holographic vector-field electron tomography to reconstruct the remanent magnetic states in CoNi nanowires with 10 nm resolution in 3D, with a particular focus on domain walls between remanent states and ubiquitous real-structure effects stemming from irregular morphology and anisotropy variations. By tuning the applied magnetic field direction, both longitudinal and transverse multi-vortex states of different chiralities and peculiar 3D features such as shifted vortex cores are stabilized. The chiral domain wall between the longitudinal vortices of opposite chiralities exhibits a complex 3D shape characterized by a push out of the central vortex line and a gain in exchange and anisotropy energy. A similar complex 3D texture, including bent vortex lines, forms at the domain boundary between transverse-vortex states and longitudinal configurations. Micromagnetic simulations allow an understanding of the origin of the observed complex magnetic states.

## 1. INTRODUCTION

Ferromagnetic nanostructures have been vastly studied over the past two decades, motivated by the prospect of developing improved spintronic devices based on the physics of magnetic soliton (e.g., domain walls, bubbles, skyrmions) dynamics.[1–3] Advances in nanofabrication[4–7] and measurement tools and techniques[8–10] have not only enabled additional miniaturization to improve the power and efficiency of devices but further lead to the exploration of more sophisticated three-dimensional (3D) magnetic nanostructures as candidates for new types of data storage and logic devices.[11–13] In particular, geometrical confinement on length scales comparable to the characteristic magnetic length scales has proven to be a powerful concept to stabilize 3D magnetic textures with unique properties that would otherwise not form [14,15]. The complexity of these magnetic configurations relies on the competition between the local exchange, anisotropies, and long-range dipole-dipole couplings enforcing magnetic flux closure.[16] Moreover, localized solitonic configurations of the magnetization in a nanoscale specimen, such as multifarious magnetic domain walls (DWs), generically break spatial parities present in the material. This magnetic chirality has important implications for the possibility to exert torques and to drive such configurations, e.g., by electrical currents or by external fields – one of the basic operations of spintronics. Detailed control of these specific magnetization configurations is therefore crucial for technological advances in nanomagnetism. That particularly also includes consideration of "real-structure" effects from, e.g., granular materials, misalignment of effective anisotropy axes, inhomogeneous surface anisotropies, morphology variations, or local imperfections and defects, which are known to exert a large influence on magnetic configurations and their properties, e.g., DW velocities,[17,18] in practically all realistic materials systems.

Consequently, even geometrically simple cylindrical magnetic nanowires (NWs) of magnetic materials with strong perpendicular magnetocrystalline anisotropy have been shown to exhibit a rich phase diagram of magnetic configurations ranging from uniform magnetization to longitudinal and transverse (multi-)domain states, depending on diameter, length, aspect ratio, externally applied field, and magnetic history of the NW.[19–24] Cylindrical magnetic CoNi NWs with a hexagonal close-packed (*hcp*) crystal structure oriented with their *c*-axis close to perpendicular to the NW axis are a prominent member of this class, which may be additionally tuned such to also display face-centered cubic (*fcc*) crystal structure with weaker crystalline anisotropy.[25–29] Previous experimental studies have revealed the existence of (A) a remanent longitudinal vortex state stabilized through the application of a longitudinal external field and/or moderate Ni-to-Co content ratio, and, more recently, (B) a multi-vortex/multi-transverse domain state stabilized by the application of a transverse field and/or low Ni-to-Co content ratio.[30,31] It has been revealed that the longitudinal vortex core is displaced off-center in response to crystalline anisotropy and that particular boundary regions emerge between longitudinal vortex states of opposite chirality and longitudinal and transverse-vortex states.

Notwithstanding the identification of these general states and their fundamental properties, a detailed understanding of the important transition regions between the different domains, referred to as domain walls in a generalized sense in the following, is largely missing.[32,33] A reason for this is a lack of experimental techniques allowing to probe these complex 3D nanomagnetic configurations with strong hysteretic character down to nanometer resolution and the elusive exploration of solitonic states with solely theoretic tools (micromagnetic simulations) in the presence of unknown parameters such as effective anisotropies, real-structure effects and unknown details of the magnetic history. Consequently, questions concerning the precise magnetic configuration and energetics (e.g., the balance between reduced anisotropy and increased demagnetizing field) of the DW between longitudinal vortices of opposite chirality and between regions with different crystalline texture, but also the general impact of morphology, effective magnetocrystalline anisotropies, or defects remain largely unresolved in these systems to date. Similarly, 3D effects in the transverse multiple vortex state (e.g., induced by the curvature of the NW) have not been addressed previously for the same reasons.

In the following, we will address these points by combining the first 3D reconstruction of pertaining magnetic configurations in Co/Ni NWs at nanometers resolution with detailed micromagnetic simulations considering the experimental geometry. The step from 2D to 3D magnetic imaging represents a huge methodological challenge for the pertaining electron microscopy and x-ray imaging techniques, which have been overcome only recently.[34–40] Of those, holographic vector-field electron tomography (VFET) in the transmission electron microscope (TEM) currently provides for the largest spatial resolution slightly below 10 nm, which has been demonstrated in a recent reconstruction of all three components of the magnetic induction in a cylindrical NW.[41] As a high spatial resolution is paramount in this study, we adapted and employed VFET despite rather strict sample criteria, such as restrictions on the sample thickness imposed by TEM. Once the sample is prepared, the TEM instrumentation offers a large range of characterization modes exploiting the manifold electron-specimen interaction, allowing different kinds of measurements to be performed on the same sample region, including local structural (e.g., by nano diffraction) and compositional analysis (e.g., by electron energy loss spectroscopy).

In the following, we use the above-mentioned methodology to image the nanoscale material basis and the magnetic configuration in a Co-based ferromagnetic cylindrical NW with a diameter of 70 nm displaying a dominant transverse magnetocrystalline anisotropy, with the *hcp c*-axis oriented almost perpendicular to the NW axis, and a definite granular crystalline microstructure. We employ holographic VFET to explore the 3D magnetic remanent state (RS) of the CoNi NW, where two characteristic different RSs are distinguished after applying an external saturation field of 2 T either (i) perpendicular or (ii) parallel to the NW axis, referred to as RS-1 and RS-2, in the following. The experimental results are combined with numerical simulations of co-existing magnetic states by micromagnetic continuum theory to resolve their complex nature. By microscopically resolving the magnetic configurations for two RS in such wires, we show that these states resemble earlier suggestions regarding possible ground- and metastable states but are considerably more complex, particularly concerning the DWs separating the domains.

## 2. RESULTS AND DISCUSSION

Figure 1 shows the morphology and structure of a 1.6 µm long segment of a representative cylindrical $Co_{85}Ni_{15}$ NW as determined by a combination of several TEM-based techniques. The 3D electrostatic potential reconstructed from holographic VFET (see experimental details), displayed in Figure 1(a), reveals a small deformation of the NW with respect to an ideal cylindrical geometry, where the NW is fairly straight, but with a slightly oval-shaped and irregular cross-section in different regions of the wire, and a split in the NW tip. The bright-field (BF) TEM image in Figure 1(b) reveals a projected NW width of 70 ± 5 nm in the imaged region and the existence of differently oriented crystallographic grains visible due to change in diffraction contrast. A detailed structural analysis (Figs. 1(c)-1(f)) was performed using the commercial ASTAR software environment, a scanning precession electron diffraction (SPED) assisted automated crystal orientation mapping (ACOM-TEM) (see experimental details). The SPED results show the coexistence of *hcp* (red) and *fcc* (blue) crystal phases within the studied NW segment: a central long *hcp* grain surrounded by regions with a polycrystalline texture of a mix between *fcc* and *hcp* grains. The *hcp* grain extending ~1 µm along the NW axis is oriented with its *c*-axis (magnetocrystalline easy axis) tilted at a mean angle of 75° to the NW axis, consistent with a previous report.[22] This *hcp* grain is the region of interest for the following experiments. The above variations in morphology and effective magnetocrystalline anisotropy arising from different grain phases and orientations are common for such nanowires synthesized by template-assisted electrodeposition.[23,42] In combination with inhomogeneous surface anisotropies and crystal defects, they impose important restrictions on the magnetic configuration and their dynamics, which we refer to as real-structure effects in the following.

Figs. 1(g) and 1(h) show magnetic phase images of two principal remanent magnetic configurations, where two different RSs in the *hcp* grain can be stabilized by varying the external saturation field direction: a transverse-vortex chain[22] with a perpendicular $H_{sat}$ (RS-1) and longitudinal states when a parallel $H_{sat}$ is applied (RS-2). Moreover, we observe domain boundaries between different longitudinal states and at boundaries between *fcc* and *hcp* grains. Note, however, that the details of the magnetic configuration, particularly at the domain boundaries, cannot be recovered from such 2D projections. To recover their full 3D texture, a tomographic reconstruction from multiple projections is required, as detailed below.

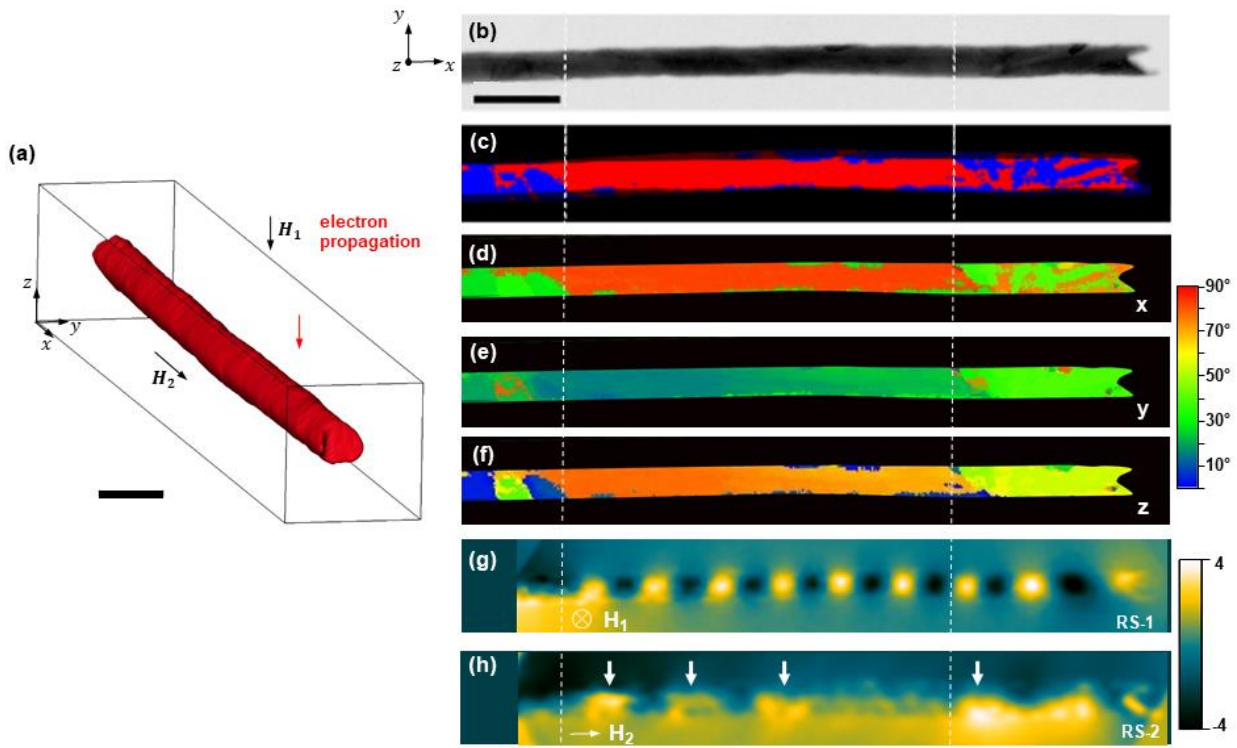

*Figure 1:* Morphology, crystallographic structure, and magnetic phase shift of a typical CoNi NW sample after applying an external magnetic field in the direction perpendicular ($H_1$) and parallel ($H_2$) to the NW axis. (**a**) Iso-surface

*rendering of the CoNi NW extracted from the 3D electrostatic potential reconstructed by electron holographic tomography, where the scale bar marks 100 nm. **(b)** Bright-field (BF) TEM image of the region of interest. **(c)** Structural map obtained by SPED ACOM-TEM, where blue and red color marks the highest match for respectively fcc and hcp crystal phases. **(d)-(f)** Crystal grain orientation maps represented by the color bar indicating the degrees of deviation of $0001_{hcp}$ and $111_{fcc}$ crystal orientation with respect to the x-, y-, and z-axes. **(g)-(h)** Magnetic phase shift along the electron propagation direction as indicated in (a) reconstructed by electron holography for the two remanent states: RS-1 (**$H_1$**) and RS-2 (**$H_2$**) with the color scale from -4 to +4 radians. The scale bar in (a) is common to (b) and equal to 200 nm. Images from (b) to (h) have the same scale.*

## A. Micromagnetic simulations of idealized case

To facilitate a comprehensive discussion of observed 3D RSs, particularly including domain boundaries, we begin with micromagnetic simulations using the OOMMF code package.[43] These preliminary micromagnetic simulations were performed for an idealized, perfectly cylindrical CoNi NW consisting of three segments (*fcc-hcp-fcc* grains) with dimensions and morphology schematically described in Figure 2(a). Although the simulations were performed for an idealized version of the studied CoNi NWs, they represent the experimentally observed magnetic configurations in their basic features. Consequently, they are well suited as a starting reference to explain the origin and for discussing the following experimental results, which are additionally affected by a host of real-structure and hysteretic effects. Figs. 2(b) and 2(c) depict 3D magnetization maps of the RSs after applying a saturation field of 2 T perpendicular (along positive *z*-axis) and Figs. 2(d)-2(g) parallel (along the positive *x*-axis) to the NW axis to mimic the magnetic history of the experimentally observed RSs in Figs. 1(g) and 1(h). In the *fcc* grains, a parallel magnetization alignment is created in both cases, with a magnetization flux closure at the NW ends to minimize the magnetostatic stray fields energy.[44,45] As in the experiments, the direction of the applied field determines the remanent magnetization configuration of the *hcp* grain, where again two types of magnetic states are found: (RS-1) a transverse-vortex chain when a perpendicular $H_{sat}$ (see Figure 2(b)), and (RS-2) two longitudinal vortex states separated by a DW when a parallel $H_{sat}$ is applied (see Figure 2(d)). These two 3D magnetic states have been observed previously and predicted to have similar energies[46] in cylindrical NWs with a diameter of around 80 nm.

In RS-1, the $H_{sat}$ applied perpendicular to the NW axis induced the formation of a chain of transverse-vortex states with alternating helicity.[22,29,46] The vortex cores are aligned close to the middle of the NW and have the same polarity with cores pointing in the positive *z*-direction, *i.e.*, $H_{sat}$ direction (see Figure 2(c)). The vortices have a core-to-core distance of about 75 nm. While these characteristics are true for the transverse-vortex states in the middle of the chain, the ones near the *fcc* grains have tilted vortex cores and even opposite polarity (Fig. 2(c)). The same figure also shows that the vortices approaching the *fcc|hcp* domain boundary do not have their cores aligned at the center of the NW axis, so their magnetization rotation is not symmetrical as compared to the rest of the vortices in the middle of the chain. This corresponds with a similar transverse-vortex chain configuration in a study comparing micromagnetic simulations with 2D magnetic imaging.[22]

RS-2 (Figure 2(e)) consists of two longitudinal vortex states[32,47] with opposite chirality but the same polarity, where the cores are aligned towards the $H_{sat}$ direction (positive *x*-axis) (see Figure 2(e) and 2(g)). The presence of longitudinal vortices at remanence (also called vortex structure) has previously been observed in single-crystal and textured cylindrical Co-based NWs with a uniaxial anisotropy.[28,32] The magnetization vectors generally have an *x*-component predominantly pointing to the previously applied saturation field direction, revealing a slight canting of the vortex states. This canting effect of the longitudinal vortex states is caused by the tilted magnetocrystalline anisotropy, which favors an alignment of magnetization vectors with the nearest easy axis direction (here, at 77° relative negative *x*-axis, see Suppl. Material note 1). For the same reason, the vortex cores are slightly shifted off-center.

The longitudinal vortex states are separated by a generalized DW, where individual magnetization vectors rotate through a transverse state while going from left to right vortex state with opposite chirality. However, there is no switch of the magnetization direction of the longitudinal vortex domains relative to the NW axis, *i.e.*, no head-to-head or tail-to-tail polarity domains like in a transverse vortex DW. We, therefore, call this a chiral DW (CDW), referencing the change in the magnetization rotation around the wire axis without a switch in the polarization of the domains. At the middle of the CDW, the magnetization vectors are oriented parallel to each other (Figure 2(g)) at an average angle of 50° with respect to the NW axis, which approaches the direction of the magnetocrystalline easy axis. These DWs were predicted by micromagnetic simulations over a decade ago[48,49] and experimentally observed in a publication by Y. P. Ivanov et al.,[33] however, a thorough discussion of the detailed structure or energetics (CDW energy) has not been conducted yet.

These idealized simulations serve as a reference for the configuration in a real NW as obtained by VFET. The experimental results contain distinct structural and morphological features of the NW and are affected by persisting limitations of the holographic VFET in terms of spatial resolution, incomplete tilt range, and noise. To perform a simulation closely resembling the experimental case, we conducted an additional set of micromagnetic simulations for a pure *hcp* crystal structure, using the same magnetic parameters but dimensions and morphology of the real NW volume as found experimentally from the 3D electrostatic potential reconstructed by VFET. These simulations are in astonishing agreement with the remanent magnetic states observed experimentally, despite not considering local changes in crystal orientations and grain boundaries, which demonstrates the crucial impact of the NW morphology details.

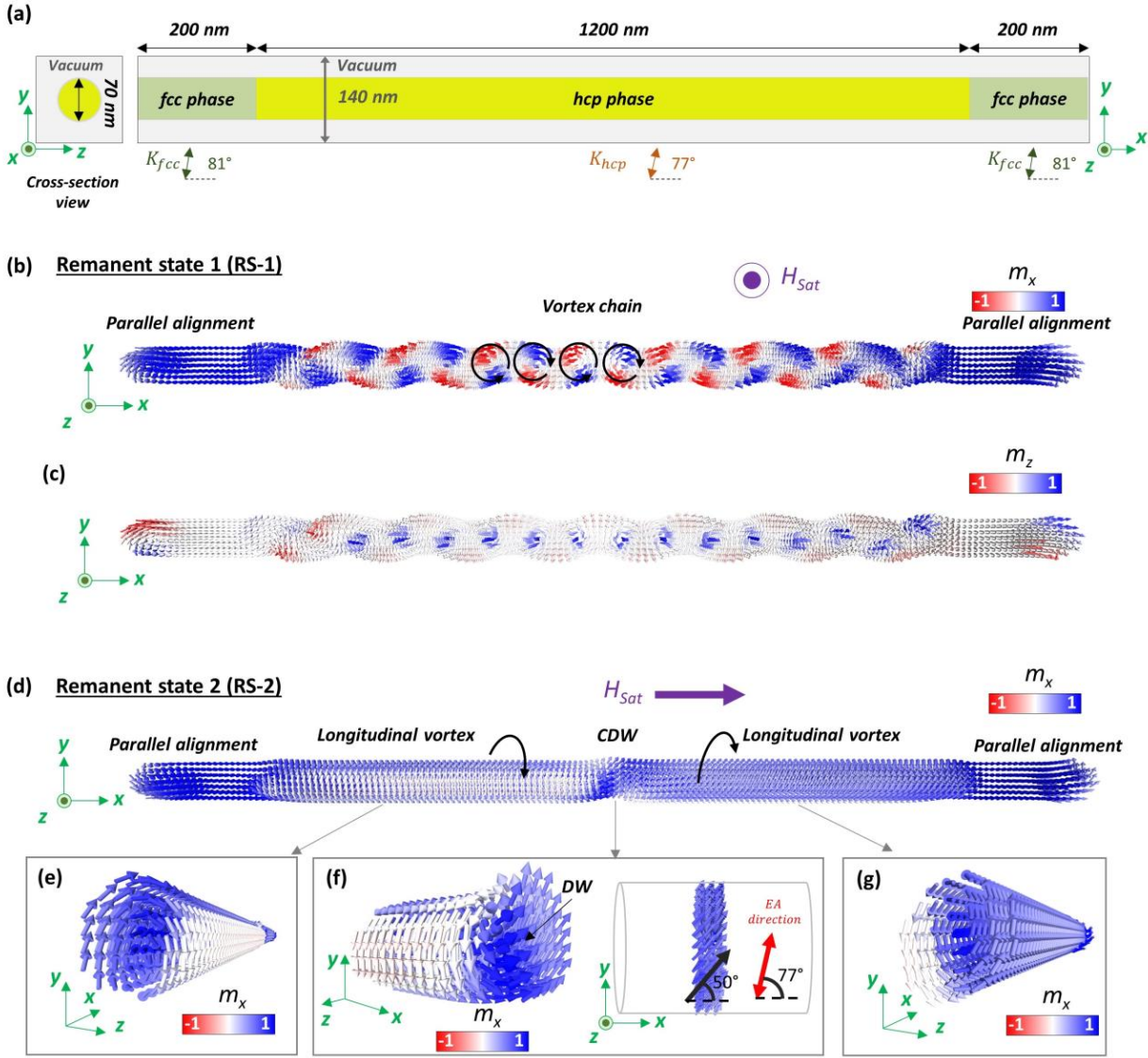

*Figure 2:* 3D magnetization of a simplified/idealized CoNi NW computed by micromagnetic simulations. (a) Scheme of the geometry and crystal structure of the simulated CoNi NW. (b,c) 3D vector maps of the RS after the application of a saturation field ($H_{sat}$) perpendicular to the NW axis with color-coding of (b) $m_x$ and (c) $m_z$, the reduced components (normalized with the saturation magnetization). (d) 3D vector map of RS after the application of $H_{sat}$ parallel to the NW axis with color-coded $m_x$. (e-g) Cross-section views of longitudinal vortex states with (e) positive and (g) negative chirality, and (f) the chiral domain wall (CDW) in between.

**B. Perpendicular saturation: transverse-vortex chain**

We begin the discussion of the tomographically reconstructed magnetic induction vector-field with the RS-1 configuration after applying a 2 T saturating magnetic field perpendicular to the NW axis along the *z*-axis (*i.e.,* also almost perpendicular to the crystallographic *c*-axis). Figure 3 shows the main results compared to realistic micromagnetic simulations based on the real morphology obtained from the experimental tomogram (see Figure 3(c) as well as 3D iso-surface rendering in Suppl. Material Movie 1). The reconstructed magnetic stray field around the NW has been removed because it is influenced by experimental artifacts due to its weaker signal-to-noise ratio compared to the magnetic flux density inside the NW. In the *xy*-plane (Fig. 3(a)), a series of transversal vortices with alternating direction of rotation (chirality) is observed. This coincides with previous findings,[22] where the magnetic configuration of the same type of NWs was studied by 2D electron holography. The *xz*-plane (Fig. 3b), which contains the direction parallel to the electron beam, reveals homogeneous alternating segments of positive (blue) and negative (yellow) *y*-components of the **B**-field, suggesting this transverse magnetic texture to be virtually invariant in the *z*-direction (see also 3D volume rendering in Suppl. Material Movies 2,3). Accordingly, in Figure 3(b), the **B**-field obtained from the micromagnetic simulation (corresponding magnetization in Figure 2(b)) is visualized using the same color scale and positions as the experimental 3D data in Figure 3(a). The similarities between experiment and simulation are striking regarding their overall appearance in terms of size, e.g., 75 nm average distance between vortices, and shape. A detailed view inside a single vortex, corresponding to the red and blue rectangles on experimental and simulated results, respectively, is shown in Figs. 3(d)-3(f) for all three planes *xy*, *xz*, and *yz*. While the experimental data allows for resolving the magnetic *z*-component in a few vortex cores only, they suggest an overall core alignment in the same direction due to the applied saturation field. Based on micromagnetic simulations, such a spin configuration is characteristic for all the vortex cores along the entire vortex chain, with the core pointing in the same (positive) *z*-direction but with a slight continuous transition in the axial direction toward the *end of the chain.* Consequently, all the vortices in the middle of the chain favor an in-plane (*xy*-plane) rotation of the magnetic vectors around a core oriented along the *z*-axis (out-of-plane), and therefore perpendicular to both the NW axis and the alternating transversal magnetic domains.

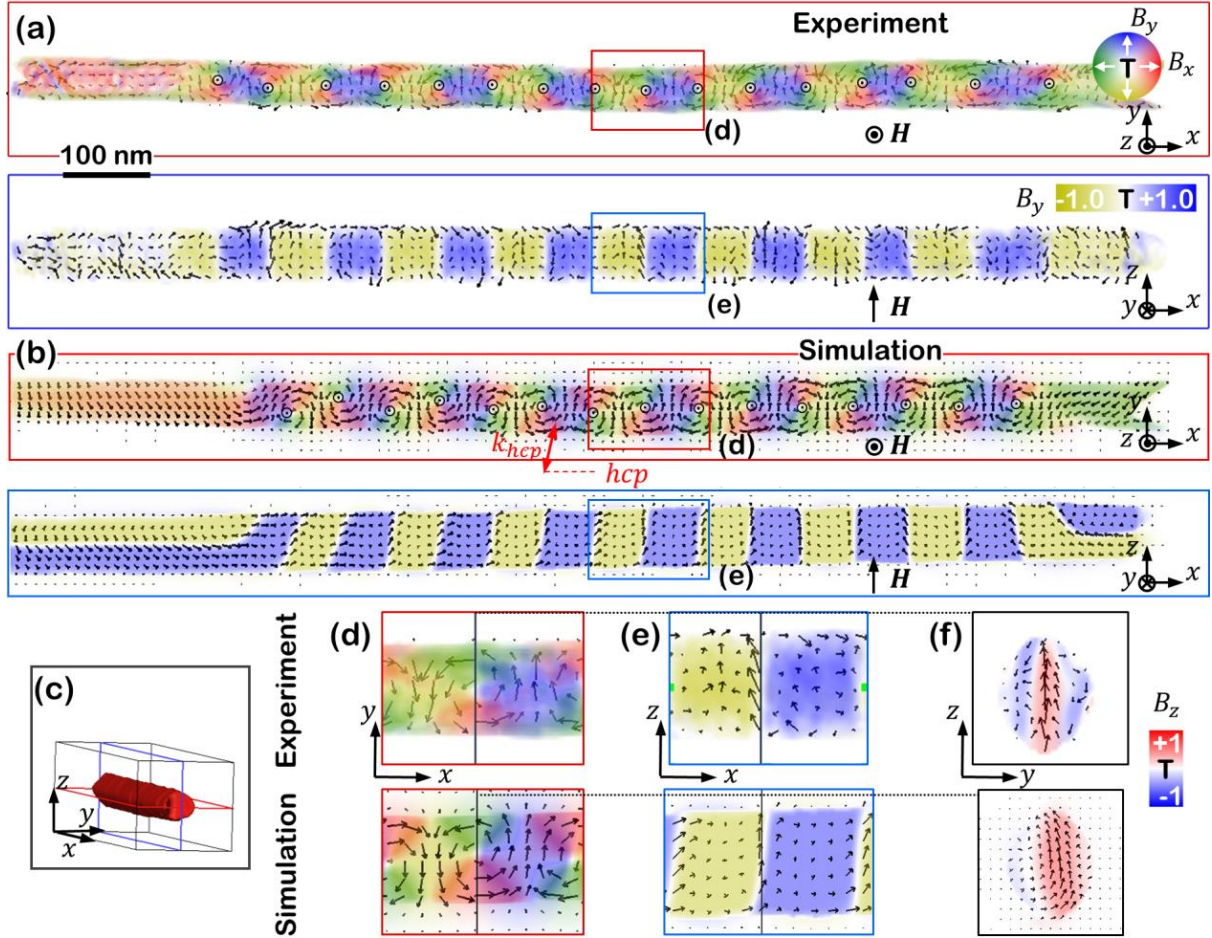

*Figure 3:* *3D reconstruction of the magnetic **B**-field inside a CoNi NW by holographic VFET and comparison with micromagnetic simulation after applying a saturation field (**H**) in the z-direction, i.e., perpendicular to the NW axis (indicated by black arrow). (a) Central slices in axial (xy) and (xz) direction through the 3D **B**-field inside the NW visualized by arrow-plots and color-coded volume rendering of $B_x$ and $B_y$ in Tesla (T) revealing a transverse vortex chain in the hcp region. (b) **B**-field from the micromagnetic simulation of the NW incorporating the real NW morphology and assuming hcp-single-crystal wire. (c) Iso-surface rendering of the electrostatic potential showing the NW morphology. The positions of the axial slices in (c) are depicted as red and blue rectangles in (a) and (b). (d,e) Zoom-in of the regions indicated in (a) and (b). (f) Cross-sectional slice (yz) through the vortex-core marked as a black vertical line in (d,e) superimposed by the color-coded $B_z$ component. The scale bar in (a) is common to all images in (a) and (c) and is equal to 100 nm.*

The overall transverse vortex chain configuration can be compared to a double Halbach configuration,[50] where the driving effect for the chain formation is the minimization of stray fields outside the structure by concentrating the fields within the NW. Accordingly, the configuration can be considered a string of antiparallel perpendicular domains (aligned with the *y*-axis), separated by transverse DWs, *i.e.,* the vortices pointing in the *z*-direction. Furthermore, there is an almost complete absence of variations of the observed texture along the *z*-axis in the middle of the vortex chain.

Apart from the transverse vortex configuration within the chain, there is also a transition into an axial configuration at the ends of the chain. Focusing on the region at the left side of the transverse-vortex chain, several particular features can be distinguished from the 3D data. First of all, the magnetic configuration starts to evolve from a transverse vortex chain into a longitudinal vortex. This transition, which may also be considered a domain wall between transverse and longitudinal vortex state, is fully established in the simulation, whereas the experiment shows are more complex behavior (see further below). When approaching the domain boundary interface, the center of the transverse vortices (marked in Figs. 3(a) and 3(b) as black circles and blue/red vertical lines) deviate from the NW axis in a characteristic intensifying zig-zag pattern. Additionally, the vortex lines are tilted with respect to the $z$-axis. A possible explanation of this peculiar phenomenon is the accommodation of a continuous transition to the adjacent longitudinal configuration. In the simulations this change is believed to come from changes in the NW shape, and the transition is from transverse-vortex states into a longitudinal vortex domain. The topologically protected longitudinal vortex core (indicated by a red line in Fig. 3(b) and also discussed in Fig. 5(f) and 5(g)) must be expulsed from the NW, thereby distorting and shifting neighboring vortex cores forming the vortex chain. This mechanism breaks the symmetry of the vortex chain arrangement and promotes a shift of the vortex cores away from the wire's $xz$-mid-plane. A similar mechanism characterized by a bending of vortex lines is also observed in the longitudinal vortex configuration, as discussed further below. This is believed to be the origin of the transition between longitudinal and transverse vortex states, where the vortex line in a longitudinal vortex bends out towards the surface of the nanowire, transitioning into transverse vortex states. An example of such a transition observed by 2D electron holography is presented in Suppl. Material note 5. While a similar tilting of the vortex lines is also observed in the 3D experimental results, the resulting longitudinal domain is more turbulent, i.e., neither a homogeneous parallel domain, nor a clear longitudinal vortex. This is believed to be due to the multi-granular structure containing an increasing amount of *fcc* structure that proceeds the magnetic transition region. This is in agreement with previous results showing variations in the curling component of longitudinal vortex domains in NWs with a mixture of *fcc* and *hcp* crystal phase as discussed in a previously mentioned paper (Andersen et al.[22]).

Both the simulated and the experimental vortex chain also show a slight variation in the core-to-core distance, where the last couple of vortices before the right-hand NW end are slightly more spread out. Moreover, the vortex chain in the experimental results contains one extra vortex core (15) compared to the simulated (14), and its effective chain region stretches over a longer part of the NW. While the general inter-core distance is similar for the two cases, there is a difference when analyzing the region before the NW tip. The simulated vortex chain only displays a slight increase in distance between two vortex cores towards the end of the NW (around 10% between smallest and largest spacing), whereas the experimental configuration shows an inter-core distance of the last two vortices before the NW tip (rightmost in Figure 3(a)) being more than 40% of the mean spacing. One possible explanation for this discrepancy is the fact that the simulation does not consider local variations in crystal grains and orientations of the real NW. By correlating the magnetic configuration (Figure 3(a)) with the crystal structure (Figures 1(c)-1(f)), the last vortex is located in a polycrystalline region consisting of several crystal grains. In addition, the proximity to the NW tip and the vortex chain makes it energetically more favorable for the chain to continue past the large *hcp* grain. Also, the relative +77° orientation of the *c*-axis with respect to the positive or negative *x*-axis (two degrees higher than the average measured value) is no longer trivial for the resulting simulated RS (Suppl. Material note 1,2).

## C. Parallel saturation: longitudinal vortex domains

Figure 4 presents the main results of holographic VFET and micromagnetic simulations performed at remanence after applying the magnetic saturation field along the NW axis (RS-2). According to micromagnetic simulations performed in the idealized cylindrical NW, a similar configuration is expected to be established even if the applied saturation field is oriented with only a few degrees of variations from the perpendicular direction (see Suppl. Material note 3). This is important, as the experimental procedure does not allow for applying the external magnetic field perfectly parallel to the NW axis (see Experimental details). In Figure 4(a), the resulting magnetic *B*-field is visualized for both the *xy*- and *xz*-plane (see also 3D volume rendering in Suppl. Material Movies 2,3). As in the case of RS-1, the magnetic stray fields have been removed from the tomogram. In both experiment (Figure 4(a)) and micromagnetic simulation (Figure 4(b)), the CDWs are present in crossover regions indicated by a blue-white-yellow color map. The adjacent longitudinal vortex domains on both sides of the CDW are well-represented in the corresponding *xz*-planes. All vortex domains have a positive *x*-component of the *B*-field, indicating equal polarity but different chirality. This is particularly visible in the detailed view of the CDWs (Figures 4(c)-4(f)), which also clearly exhibit the left- and right-handed vortex lines being expulsed to the NW surfaces because they cannot be transformed into each other in a continuous way (see also Fig. 5). The different rotation of the vortex domains can be clearly seen in the cross-sectional views in the *yz*-plane for both experimental (Figure 4(g)) and simulated (Figure 4(h)) results. The longitudinal vortices are slightly canted with respect to the NW axis, producing a small number of magnetic charges at the NW surface and stray fields (Figure 1h) outside the structure, not only at the CDWs. The canting is also linked to the magnetocrystalline anisotropy easy axis of the *hcp* grain.

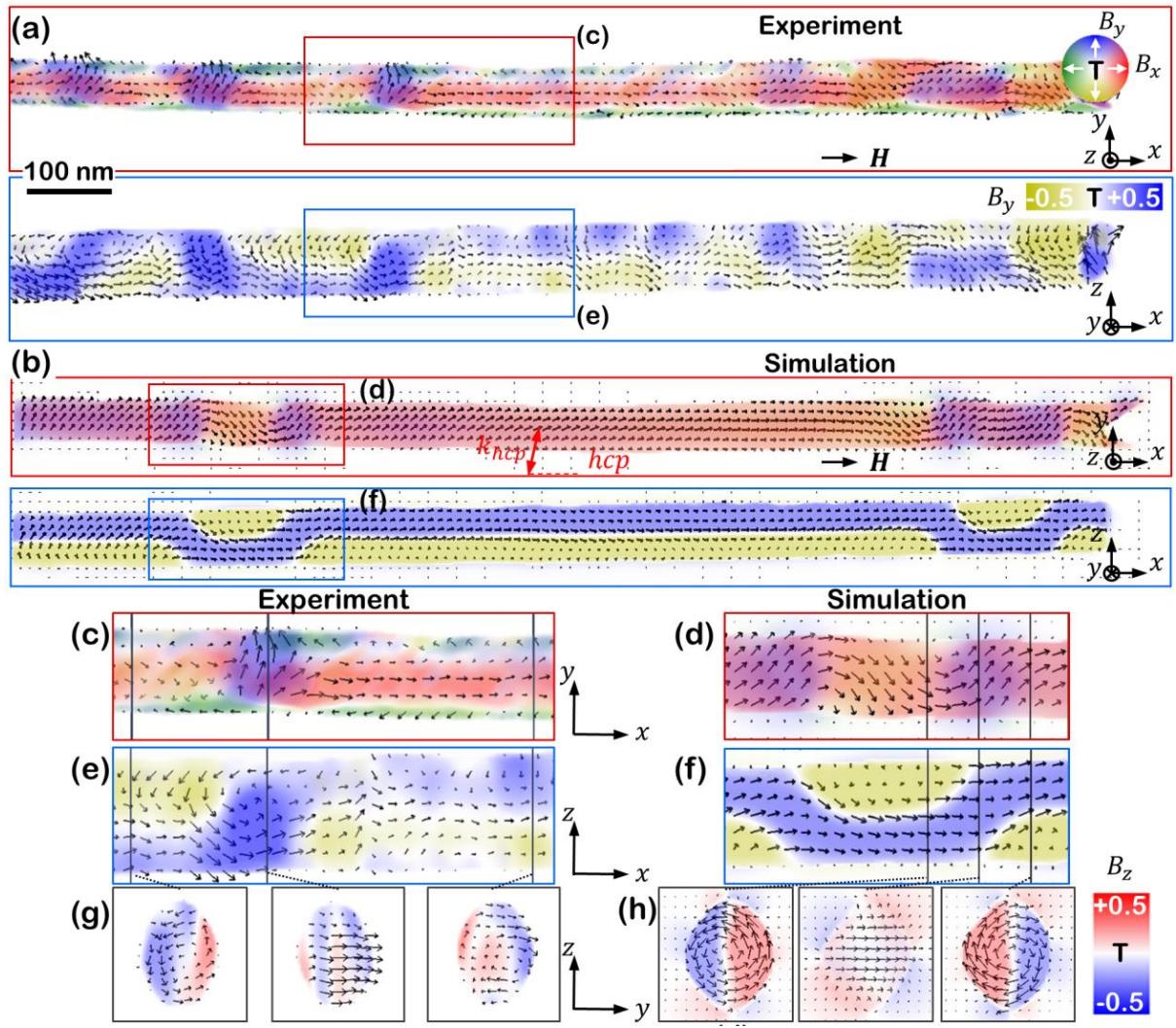

*Figure 4. 3D reconstruction of the magnetic **B**-field inside a CoNi NW by holographic VFET and comparison with micromagnetic simulation after applying a saturation field (**H**) in the x-direction (indicated by black arrow), i.e., parallel to the NW axis. (a) Central slices in axial (xy) and (xz) direction through the 3D **B**-field inside the NW (axial slices indicated in Figure 3c) visualized by arrow-plots and color-coded volume rendering of $B_x$ and $B_y$ in Tesla (T). (b) **B**-field from micromagnetic simulation incorporating the real NW morphology and assuming hcp crystalline anisotropy (parameter $k_{hcp}$). The simulation results are of the same cross-sectional slices and visualized in the same way as the experimental data. (c-f) Zoom-in of the regions indicated in (a) and (b). (g,h) Cross-sectional slices (yz) at positions marked as black vertical lines in (c-f) superimposed by the color-coded $B_z$ component show the transition between two vortex states with opposite chirality separated by a CDW. The scale bar in (a) is common to all images in (a) and (b) and is equal to 100 nm.*

Again, there are certain differences between the experiment and the real-shape simulation in terms of the number of CDWs and their positions, e.g., the obvious existence of an additional CDW in the experiment (Figure 4(a) and 4(b)). Furthermore, the orientations of the CDWs are more perpendicular to the NW axis in the experiment compared to the simulation. Since the observed NW is not a monocrystalline homogeneous cylinder (Figure 1a-b), the DW positioning might be affected by pinning sites originating from modulations in the NW thickness or crystal phase. Conclusively, it is more energetically favorable for the system to form multiple domains, where the DW position is influenced by a combination of the NW morphology, as well as the position and orientations of crystal grains, thus explaining the difference in location and number of DWs between the simulated and experimental RS.

**D. Domain boundary / domain wall energetics**

Having completed the detailed description of the magnetization textures in the observed remanent states RS-1 (transverse-vortex chain) and RS-2 (longitudinal vortex) focusing on the domain boundaries/walls occurring in these systems, we finally investigate the formation energies of the latter. Fig. 5(a) displays the vortex line skeleton (determined by tracing the **B**-field vorticity maxima) at the boundary between longitudinal and transverse-vortex domain (see also Fig. 3(b), 4(b), and Suppl. Material Movies 4 and 5) together with the corresponding 1D micromagnetic energy densities, obtained by integrating the volume energy densities obtained from the micromagnetic simulations over the *yz* cross sections. While this particular domain wall is not present in the 3D reconstructed NW region because of the magnetization history and the mixed *fcc* and *hcp* phase at the left NW tail, there is plenty of experimental evidence of such a transition from other CoNi NW (regions). An exemplary combined holography and micromagnetic study of such a pure *hcp* region, which was not deliberately polarized by an external field, is presented in Suppl. Material note 5.

In Fig. 5(a), one clearly observes the reduction of the demagnetizing field energy in the transverse vortex cores, which is compensated by an increase in exchange and anisotropy energy. The comparison of average total energies between longitudinal vortex state region and the transverse-vortex chain state shows a slightly higher stability of the former, which explains why the transverse-vortex chain is only formed when the external field is perpendicular to the NW and the crystallographic *c*-axis. The domain wall separating both regions is comparable in total energy to the transverse vortex chain state. Comparing the domain wall (transition) between longitudinal and transverse vortex state and the CDW in Fig. 5(b) we furthermore note that both a related, i.e., exhibit a similar bending of the vortex line towards the longitudinal vortex.

The CDW formation relies on a decrease of both exchange and anisotropy energy while increasing the demagnetizing field energy compared to a single longitudinal vortex (Fig. 5b). The reduction of exchange energy occurs due to the suppression of the vortex texture in the CDW. This promotes a parallel alignment of the magnetization vector, as shown in Fig. 2(f)). The anisotropy energy is then reduced by the alignment of the CDW magnetization with the crystallographic *c*-axis. Note that the increase of the demagnetizing field energy surmounts the combined exchange and anisotropy energy drop, rendering the single CDW configuration metastable. However, the RS of both real-shape simulation (Figure 4b) and experimental results (Figure 4a) contains more than two longitudinal vortex domains separated by multiple CDW. This configuration decreases the demagnetizing field energy, which helps to stabilize the observed multi-CDW state during creation (see also Ref. 33 for other multi-CDW states). The observed energetics of the CDW sets it apart from conventional DW between homogeneously magnetized domains as a rather unusual kink or exchange and anisotropy energy gaining texture, with unique static (i.e., stray field generating) and dynamical properties.[32]

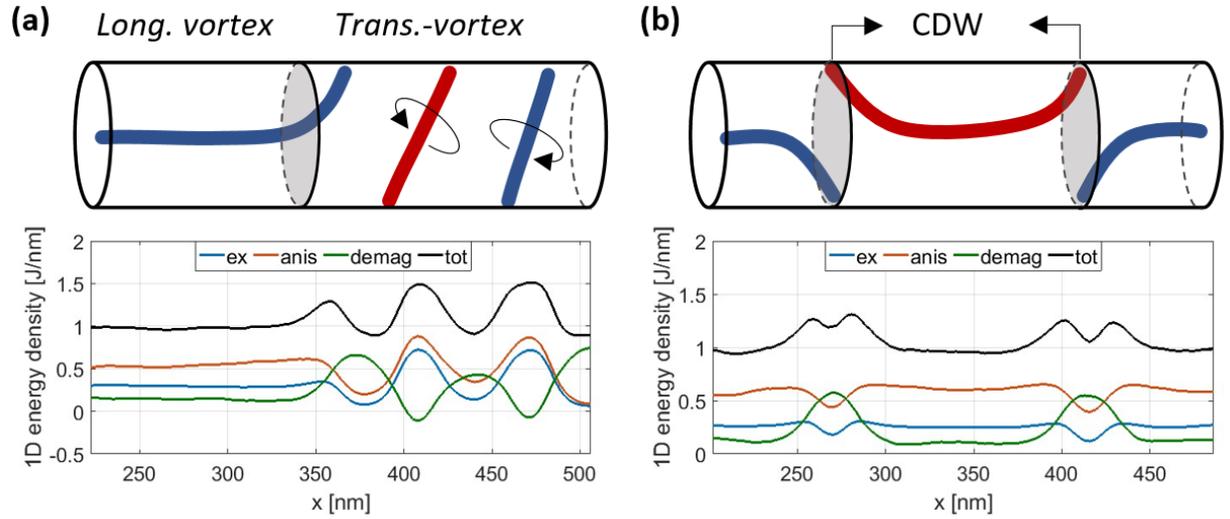

*Figure 5. Vortex line skeleton and 1D energy densities of (a) longitudinal and transverse-vortex domain boundary and (b) CDW. The blue and red lines indicate the position of the clockwise and counterclockwise rotating vortex lines as obtained from micromagnetic simulations.*

## 3. SUMMARY

We have conducted a comprehensive 3D investigation of the magnetic remanent states in cylindrical NW including pertaining domain walls and boundaries, tunable through the orientation of an external magnetic saturation field. To this end, we have performed 3D quantitative nanoscale magnetic imaging in Co-rich CoNi NWs with a nearly perpendicular magnetocrystalline anisotropy, revealing two distinctly different remanent configurations depending on the direction of an applied saturation field. A transverse-vortex chain is formed for a saturation field simultaneously oriented perpendicular to both the NW axis and the magneto-crystalline *c*-axis, whereas for a field applied parallel to the wire axis, a longitudinal vortex configuration develops. The transverse-vortex chain mainly displays in-plane magnetic components, with a small 3D modulation along the vortices. Characteristic zig-zag modulations of the vortex lines are observed for the transverse-vortex states located close to the chain's extremities, eventually rotating the lines into a longitudinal vortex. The second remanent magnetic configuration consists of longitudinal vortex states separated by chiral DWs, characterized by an expulsion of the opposite vortex lines from the nanowire. The chiral domain walls are furthermore distinguished by a reduction of both exchange and anisotropy energy from conventional DWs. By combining the experimental results and micromagnetic simulations using both an ideal cylindrical shape and the real-shape retrieved by VFET, we observed a large impact of real-structure effects on the resulting RSs of the system. This shows the importance of carefully controlling the system morphology for accurate manipulation of its magnetic configuration. Both magnetic configurations and their solitonic domain boundaries/walls are interesting candidates for applications such as magnetic memories. For instance, the deliberate manipulation of the topologically protected CDWs, e.g., through currents or external fields, may be exploited for racetrack memories. Similarly, deliberate manipulation of the polarity in the vortex chain may hold potentials for extremely compact magnetic memory devices.

## 4. EXPERIMENTAL DETAILS

### A. Nanowire fabrication and sample preparation

The $Co_{85}Ni_{15}$ nanowires were grown by electrodeposition into self-assembled pores of anodic aluminum oxide templates in a three-electrode cell using a Watts-type bath electrolyte. The alumina templates were made by a two-step anodization process. An electrolyte composition of 0.124 M $CoSO_4·7H_2O$ + 0.085 M $CoCl_2·6H_2O$ + 0.064 M $NiSO_4·6H_2O$ + 0.064 M $NiCl_2·6H2O$ + 0.32 M $H_3BO_3$ was used with electroplate voltage of -1.2 V versus the Ag/AgCl reference electrode, and a pH value of 3.0. To enable TEM investigation of individual wires, the electrodeposited NWs were then released from the template by chemically dissolving the alumina in a chromic oxide and phosphoric acid solution. Finally, they were transferred to a copper TEM-grid (100-mesh hexagonal) with Quantifoil S7/2 carbon support and 2 nm carbon film (Quantifoil Micro Tools, Jena, Germany).

### B. Application of the external magnetic fields

A magnetic field of 2 T into the image plane direction was applied using the objective lens of the TEM instrument prior to the measurements. The effective magnetic component relative the sample axis can then be controlled by tilting the sample plane. Where $H_{OL}$ is the amplitude of the magnetic field generated by the objective lens, we will then get that $H_1 = H$, while $H_2 = H \sin \theta$, where $\theta$ is the sample tilt. Specifically, the external magnetic field was applied at zero tilt (I), and at 70° tilt angle with tilt axis perpendicular to NW axis such that the NW tip goes into or out of the image plane. This corresponds to the application of a magnetic saturation field (I) perpendicular and (II) parallel to the nanowire axis, respectively. The magnetocrystalline easy axis has an orientation close to perpendicular to both the electron beam direction and the NW axis (Figure 1(c)-1(f)), meaning that both the applied fields were close to perpendicular to the *c*-axis.

### C. Quantitative 3D magnetic imaging

Holographic VFET was performed on the studied sample to obtain the magnetic induction of single nanowires. The holographic tilt series were recorded on an FEI Titan 80-300 Holography Special Berlin TEM instrument equipped with a 2K slow-scan CCD camera (Gatan Inc.'s Ultrascan 1000P) operated at 300 kV in image-corrected Lorentz-mode using a double biprism setup.[51] The holograms were acquired with a pixel size of 0.67 nm and an interference fringe spacing of 4.95 pixels per fringe. A dedicated dual-axis tomography sample holder (Model 2040 from E.A. Fischione Instruments, Inc.) was used for establishing high tilt angles and manual in-plane rotation as required for recording a dual-tilt-axis tomography tilt series.[41] For each remanent state of the NW (RS-1 and RS-2), a hologram tilt series were recorded around two different axes rotated by 90° in-plane with respect to each other. Then, in the case of RS-1, the sample was flipped upside down, and another two tilt series were acquired around exactly the same tilt axes as in the non-flipped case. All six series covering a tilt range from -66° to +66° with a step size of 3° were obtained by an in-house built semi-automatic software (THOMAS) that efficiently performs and monitors the tilting process and acquisition of object holograms and reference holograms.[52] Flipped and non-flipped phase image series were corrected for image distortions (see Suppl. Material note 4) and displacement, subtracted and added in order to separate the magnetic and the electric contributions to the recorded phase images retrieved from the holograms by Fourier reconstruction.[53]

From the magnetic phase tilt series, the $B_x$ (parallel component to the tilt axis) and $B_y$ (parallel component to the tilt axis after 90° in-plane rotation) components were reconstructed in 3D using weighted simultaneous iterative reconstruction technique (W-SIRT).[54] From the electric phase tilt series, the electrostatic potential $\Phi$ was reconstructed in 3D using W-SIRT. Because the experimental conditions of the VFET experiment (tilt range, tilt steps, resolution of the projection data, object size) were very similar to those reported in Ref. 41, we can assume the same spatial resolution as determined in Ref. 41, i.e., 10nm in x- and y-direction and slightly reduced resolution in z-direction due to the limited tilt range.

### D. Mapping of nanowire texture

The crystal grains and orientations of the studied nanowire were mapped using scanning precession electron diffraction (SPED)[55,56] with the commercial software package NanoMegas ASTAR (the product name), an automated crystal orientation mapping (ACOM-TEM) system.[57] The ACOM technique consists of scanning the defined sample area and acquiring PED patterns of each defined pixel in the sample scan. These patterns contain the crystallographic information of the area, allowing the user to obtain the crystal structure and orientation information, by an automatic comparison of the experimental and theoretical diffraction patterns for any given material. The mapping was done on a Philips CM20-FEG TEM operated at 200 kV with a condenser aperture of 50 µm. SPED patterns were acquired at a camera length of 235 mm and a magnification of 58 kx, scanned with an electron beam of ~1 nm spot size with a step size of 4 nm (pixel of 4 nm$^2$) over the sample and a precession angle of 0.7°.

### E. Micromagnetic simulations

The micromagnetic simulations were performed by using the OOMMF code.[58] We simulate the remanent state of an idealized 70-nm-diameter cylindrical CoNi nanowire formed by two crystal phases, and a real-shape and *hcp*-single-crystal CoNi NW. The real-shape NW morphology was obtained from iso-surface rendering extracted from the 3D electrostatic potential retrieved by holographic VFET. Imitating what was done experimentally, simulated remanent states were obtained by applying a saturation field of 2 T in two different directions: perpendicular and parallel to the NW axis. Considering that the deposition process produces CoNi NWs with possibly coexisting crystal phases (*fcc* and *hcp*),[22] the idealized NW is a textured structure formed by two 200 nm long cylindrical *fcc* grains located at each end of the wire, with a 1200 nm long cylindrical *hcp* grain in between them (Figure 2(a)). For the magnetic parameters, we consider both crystal phases to have the saturation magnetization ($M_S$) and exchange constant ($A$) of the $Co_{85}Ni_{15}$ alloy reported by Samardak et al.[23] ($M_S$ = 1273 kA/m; $A$ = 26 pJ/m); a cubic anisotropy with an anisotropy constant $K_{fcc}$ = 100 kJ/m$^3$ for the *fcc* grains;[59] and a uniaxial anisotropy with an anisotropy constant $K_{hcp}$ = 350 kJ/m$^3$ for the *hcp* grain. An easy axis orientation at 81° relative to the NW axis was chosen for the *fcc* phase considering the nanodiffraction (ASTAR) measurements performed on the studied NW and another NW containing a long *fcc* grain of a NW from the same batch as the presented NW (see Supporting Material note 1), while an easy axis orientation at 77° relative to NW axis was tuned for the *hcp* phase, for both idealized and real-shape NWs, considering ASTAR measurements and simulations at different easy axis orientations. The cylindrical and real shape of the NWs were generated by stacking magnetic unit cells of sizes 5 × 5 × 5 nm$^3$ and 2.7 × 2.7 × 2.7 nm$^3$, respectively.

## ASSOCIATED CONTENT
**Supplemental Material** accompanies this paper and contains additional details about the experimental methods, simulations, and data analysis process.

    Note 1: Easy axis direction of CoNi nanowires for micromagnetic simulations.

    Note 2: Shape effects and magnetocrystalline easy axis orientation study by micromagnetic simulations.

    Note 3: Simulated remanent states of the system with different saturation field direction.

    Note 4: Correction of anisotropic magnification distortion in phase images.

    Movie 1: CoNi nanowire tomogram of mean inner potential.

    Movie 2: CoNi nanowire tomogram of magnetic induction x-component.

    Movie 3: CoNi nanowire tomogram of magnetic induction y-component.

    Movies 4, 5: CoNi nanowire simulation of magnetic induction y-component with extracted vortex lines.

**Conflict of interest:** The authors declare no financial/commercial Conflict of Interest.

## ACKNOWLEDGEMENTS
The research leading to these results has received funding from the European Union Horizon 2020 research and innovation program under grant agreement No. 823717 – ESTEEM3. We express our gratitude to M. Lehmann (TU Berlin) for providing access to the FEI Titan 80–300 Holography Special Berlin. D.W. and A.L. acknowledge funding from the European Research Council (ERC) under the Horizon 2020 research and innovation program of the European Union (grant agreement number 715620). C.B and M. V. acknowledge funding from Spanish MINECO under Project MAT2016-76824-C3-1-R and PID2019-108075RB-C31, and the Regional Government of Madrid under Project S2018/NMT-4321 NANOMAGCOSTCM.

## AUTHOR CONTRIBUTIONS
I.M.A., D.W., L.A.R., and A.L. have contributed equally to this work. I.M.A., D.W and T.N. performed the holographic VFET experiments, and D.W aligned, reconstructed, and evaluated the 3D data. L.A.R. performed the micromagnetic simulations. D.O. and I.M.A. performed the ASTAR© measurements and the associated data treatment. C.B. fabricated the NWs. A.L. contributed with data analysis and magnetic calculations. I.M.A., A.L., D.W, and L.A.R. collaborated on the main discussions of the results and wrote the preliminary manuscript. U.K.R., M.V., C.G., and E.S. contributed to discussions and the development of the manuscript.